# Ab-initio study of short-range ordering in vanadium-based disordered rocksalt structures

Zinab Jadidi[1,2], Julia H. Yang[1,2], Tina Chen[1,2], Luis Barroso-Luque[1,2], Gerbrand Ceder[1,2]

[1] Department of Materials Science and Engineering, Berkeley CA 94720
[2] Materials Sciences Division, Lawrence Berkeley National Laboratory, Berkeley CA 94720

*Abstract* - Disordered rocksalt Li-excess (DRX) compounds are attractive new cathode materials for Li-ion batteries as they contain resource-abundant metals and do not require the use of cobalt or nickel. Understanding the delithiation process and cation short-range ordering (SRO) in DRX compounds is essential to improving these promising cathode materials. Herein, we use first-principles calculations along with the cluster-expansion approach to model the disorder in DRX Li$_{2-x}$VO$_3$, $0 < x < 1$. We discuss the SRO of Li in tetrahedral and octahedral sites, and the order in which Li delithiates and V oxidizes with respect to local environments. We reveal that the number of nearest-neighbor V dictates the order of delithiation from octahedral sites and that V are oxidized in a manner that minimizes the electrostatic interactions among V. Our results provide valuable insight for tailoring the performance of V-based DRX cathode materials in general by controlling the SRO features that reduce energy density.

## INTRODUCTION

Disordered rocksalt structures with Li-excess (DRX) are promising earth-abundant cathode materials for Li-ion batteries. (1–10) Unlike in conventional layered electrodes, there is no requirement on the long-range ordering of the cations in the rocksalt lattice, enabling the use of a variety of sustainable transition metals (TMs) and removing reliance on nickel (Ni) and cobalt (Co). (11–13) Some DRX compositions have shown specific energies as high as 1000 Wh/kg. (1,11)

Studies have shown that DRX materials contain cation short-range order (SRO), an equilibrium phenomenon occurring above the order–disorder transition temperature. (5,14,15) Size and charge affect the degree of cation SRO, which in turn controls the extent of connectivity of the Li$_4$ tetrahedra in DRX materials: A smaller high-valent TM leads to greater Li segregation into Li$_4$ tetrahedra clusters compared to a larger high-valent TM. (5) Employing divalent redox-active TMs has also been shown to promote the formation of Li$_4$ tetrahedra to a larger extent than trivalent TMs. (5)

If these Li$_4$ clusters are fully connected, a percolating network for facile Li diffusion is formed. (6,16) Controlling cation SRO is therefore important to promote Li percolation (5,14,15,17) and, subsequently, the amount of kinetically extractable Li. (5,11,14,18) The presence of SRO is not specific to DRX materials as it has been studied (19–25) in the field of metallic alloys following the work of J. M. Cowley on Cu$_3$Au in 1950. (19)

In this paper, we examine Li$_{2-x}$VO$_3$, $0 < x < 1$, a compound that has been investigated as a potential cathode material (26), to serve as a model system for DRX materials. Our goal is to analyze the delithiation process in the presence of SRO and to determine the specific SRO features that influence the voltage profile in these materials. In Li$_2$VO$_3$, Li can occupy both tetrahedral and octahedral sites, whereas V occupies the octahedral sites. We model the configuration space using a lattice cluster expansion trained on first-principles calculations with quaternary disorder (Li$^+$/V$^{4+}$/V$^{5+}$ and vacancies) on octahedral sites and binary disorder (Li$^+$ and vacancies) on tetrahedral sites. To fit the cluster expansion, we use the approach previously applied in a high-component Mn-disordered spinel systems (27–29), which is explained in more detail in **METHODS**.

In this study, we find that Li extraction preferentially begins from the octahedral environments with the largest number of nearest-neighbor (NN) V and that the number of NN V dictates the order of delithiation in the remaining octahedral sites, as sites with a higher number of NN V experience higher electrostatic repulsion. In addition, we find that at all stages of delithiation, the number of NN V$^{5+}$–V$^{5+}$ is minimized.

## METHODS

### A. First-principles density functional theory calculations

A cluster expansion (CE) is a formal representation of the dependence of the energy on the configuration of species on prescribed sites of a lattice. (30–32) The structure–energy dataset used to fit the present CE was obtained from density functional theory (DFT) (33,34) using the Vienna ab-initio simulation package (VASP) (35,36) and the projector-augmented wave (PAW) method. (37,38) For these calculations, we used the PBE exchange-correlation functional (39) with rotationally averaged Hubbard U correction, where the U value was selected based on a previously reported calibration to oxide formation energies (3.1 eV for V). (10,40) We used a plane-wave cutoff of 520 eV with reciprocal space discretization of 25 K-points per Å and electronic convergence of $10^{-5}$ eV with a convergence criterion of 0.01 eV Å$^{-1}$ for the interatomic forces for all our calculations.

B. *Cluster expansion and Monte Carlo sampling*

The CE method enables extensive configurational sampling that would not be computationally practical with DFT. The CE is written as:

$$E(\boldsymbol{\sigma}) = \sum_{\beta} V_\beta \langle \Phi_\alpha(\boldsymbol{\sigma}) \rangle_\beta, \quad (1)$$

where the occupancy string $\boldsymbol{\sigma}$ represents a particular configuration, with each element $\sigma_i$ of the string representing an occupation variable for site i. The summation in equation 1 is over all symmetrically distinct clusters $\beta$, which consist of products of point basis functions. In this work, we used the sinusoidal basis functions as described in Van de Walle et al. (41) The coefficients $V_\beta$ are effective cluster interactions (ECIs) with the multiplicity of the corresponding clusters included, (30,42) and the functions $\Phi_\alpha$ are cluster functions. (41)

The expansion represented in equation 1 can exactly capture any property of configuration when the sum is taken over all possible geometric clusters of sites. However, in practice, the expansion is truncated above some maximum-sized clusters. For our model, based on a cubic cell with a = 3.0 Å, we considered pair interactions up to 7.1 Å and triplet interactions up to 4 Å. An Ewald summation term is included in the energy model as it has been shown to increase the accuracy in CE on ionic systems. (10,43–46)

The lattice model on which a CE is defined requires specification of which species can occupy which sites. As shown in Figure 1a, Li$_2$VO$_3$ has cubic $Fm\overline{3}m$ symmetry with Li$^+$ (green) and V$^{4+}$/V$^{5+}$ (blue) cations occupying the 4a Wyckoff octahedral sites and O$^{2-}$ (red) occupying the anion sites. (26) Li$^+$ in DRX can also occupy tetrahedral 8c Wyckoff sites. (47,48) Finally, to model delithiated configurations, vacancies (shown in gray) are treated explicitly as species that can also occupy the 4a and 8c sites. In summary, our lattice model for the Li$_{2-x}$VO$_3$, 0 < x < 1, configuration space consists of {Li$^+$, V$^{4+}$, V$^{5+}$, vacancy} on the octahedral sublattice and {Li$^+$, vacancy} on the tetrahedral sublattice.

The initial set of configurations used to fit the CE consisted of Monte-Carlo-sampled configurations with low electrostatic energy using an Ewald-only Hamiltonian. (49) This initial dataset was used for the typical iterative approach of fitting, adding new low-energy Monte-Carlo-sampled structures from the fit and re-fitting with the newly generated structures until convergence. (50,51) Monte Carlo sampling is performed at varying Li compositions (Li$_{2-x}$VO$_3$, 0 < x < 1), temperatures (0 K–5000 K), and supercell sizes ranging from 3×1×1 (with 3 anion sites) to supercells of 3×4×3, 2×6×3, 2×2×9 (with 36 anion sites) to include a wide energy range of atomic configurations in the space of Li$_{2-x}$VO$_3$, 0 < x < 1. The ECIs of the converged fit are obtained by fitting to the DFT energies of 450 symmetrically distinct structures whose lattice configurations are a subset of possible charge-balanced arrangements of Li$^+$, V$^{4+}$, V$^{5+}$, and vacancies over the cation sublattice.

To build the CE, the smol python package is used. (52) We followed the sequential fitting approach for building multi-cation sublattice coupled cluster expansions previously introduced by Yang et al. (27) For training purposes, 80% of the DFT-computed structures are reserved, while the remaining 20% are utilized for testing. First, $V_0$, the coefficient of the constant basis function is set to the average energy of the training set, an assumption that holds for sampled centered basis functions. Next, the ECIs for the point-term correlation functions are determined using the lasso regularization approach (53) which is required because the charge-neutrality constraint in ionic systems reduces the dimension (or rank) of the point-term basis functions by one. (27,46) In practice, avoiding degenerate solutions in a rank-reduced space is achieved by setting an ECI for a point-term basis function to 0, which lasso regularization on the point-term correlation functions achieves. Finally, sparse group lasso is applied to solve for the ECI for the remaining correlation functions belonging to pairs and triplets. (27) The resulting fit has a 10-fold cross-validation error of 13.1 meV per site and an in-sample root-mean-squared error (RMSE) of 12.4 meV per site. Figure S1 presents a comparison of the predicted and DFT convex hulls and the ECI values of the converged fit. Based on the cross-validation error, the RMSE, and comparison between the predicted and DFT convex hulls, the CE fit appears to represent the interactions between species relatively well. We discuss some of the specific challenges of fitting this high-component system in the SI, note 1.

To construct a representative topotactic Li voltage profile of the disordered Li$_{2-x}$VO$_3$, 0 < x < 1, we use semi-grand canonical MC simulations (SGC MC) from the smol python package (52) at T = 300 K and scan the Li chemical potential, at intervals of 0.05 eV, as a function of the Li content. The Li/V$^{4+}$ metal configuration in the fully lithiated state is obtained by equilibrating at 1800 K in a 9×8×9 cell (i.e., a 648-oxygen supercell with a total of 1296 atoms) using canonical MC to simulate the as-synthesized cation ordering. The topotactic SGC MC at 300 K is then initialized from this fully lithiated configuration and calculated by freezing all V and allowing for only Li/vacancy disorder and oxidation/reduction of $V^{4+}/V^{5+}$. We equilibrate for 2 million MC proposals and sample every 2000 proposals, totaling 1 million samples per Li chemical potential.

The voltage profile is calculated from the chemical potential as (54):

$$V(x) = -\frac{\mu_{Li} - \mu_{Li}^{reference}}{Z}, \quad (2)$$

where $\mu_{Li}$ is the Li chemical potential of the cathode material; $\mu_{Li}^{reference}$ is the Li chemical potential of the anode, which we consider to be a pure BCC Li metal (−1.9 eV); and Z is the electron charge carried by one Li ion, +1e. The average voltage for the simulated profile is calculated using equation 3, with $X_1 = 2$ and $X_2 = 1$ (54):

$$\bar{V} \approx -\frac{E(Li_{X_1}VO_3) - E(Li_{X_2}VO_3) - (X_1 - X_2)E(Li)}{(X_1 - X_2)},$$
$$\text{with } X_1 > X_2, \quad (3)$$

For the SRO analysis, we used the pymatgen package (55) to track the species occupying the 12 NN edge-sharing octahedra around each octahedral site (depicted in the top panel of Figure 1b) and the 4 NN face-sharing octahedra around each tetrahedral site (depicted in the bottom panel of Figure 1b). The species that can occupy these sites include the Li in green, V in blue, and vacancies in gray, as illustrated in Figure 1b.

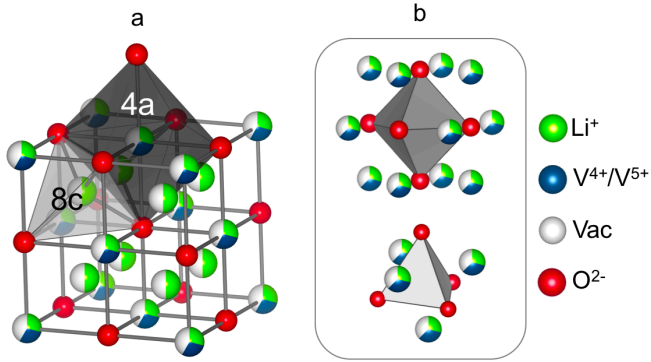

FIG. 1. (a) Rocksalt structure of $Li_{2-x}VO_3$, 0 < x < 1 with random cation occupation. The face-centered cubic (FCC) lattice of octahedral sites, which can be occupied by $Li^+$ (green), $V^{4+}/V^{5+}$ (blue), or vacancies (gray), is represented by the tri-colored spheres. The tetrahedral sites, which can be occupied by $Li^+$ (green) or vacancies (gray), are represented by the bi-colored spheres. The red spheres form the anion lattice, which is fully occupied by oxygen. (b) (top) Schematic of octahedral site surrounded by 12 edge-sharing octahedral sites; (bottom) schematic of tetrahedral site surrounded by four face-sharing octahedral sites.

## RESULTS

The SRO in the as-synthesized $Li_{2-x}VO_3$, 0 < x < 1 is analyzed in a 9×8×9 cell with the $Li/V^{4+}$ metal configuration obtained by equilibrating at 1800 K. The choice of 1800 K was made because the voltage profile derived from this configuration was found to be the most similar to the experimental voltage profile of $Li_2VO_3$ when compared to the voltage profiles obtained at lower temperatures, as depicted in Figure S2. Moreover, the simulated X-ray diffraction (XRD) pattern at this temperature (Figure S3) displays only the primary peaks associated with the rocksalt structure (38°, 43°, 63°, 76°, and 80°), suggesting that the simulation temperature exceeds the ordered semi-layered to disordered rocksalt transition temperature. Configurations sampled at temperatures below 1800 K exhibit additional peaks at low angles which have higher intensity than the main rocksalt peak at 43°.

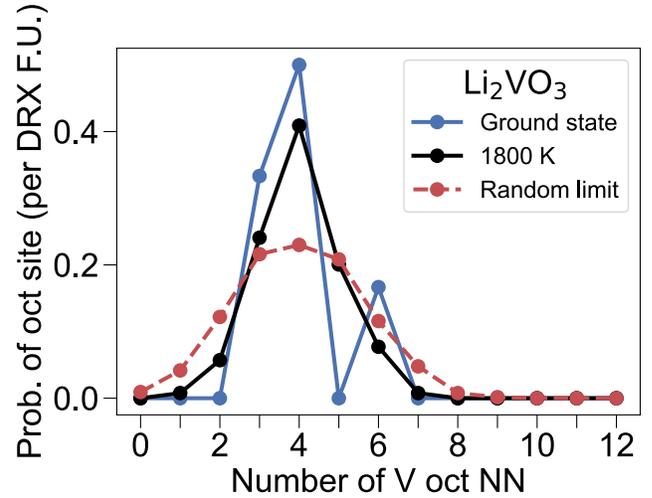

FIG. 2. Probability that an octahedron has a certain number of V-occupied NN octahedra in the ground state (solid blue curve), at 1800 K (solid black curve), and in the limit of random cation occupancy (dotted red curve). The Y axis is normalized per DRX formula unit (F.U.). Therefore, the sum of the probabilities for each curve (each T) is 1.

Figure 2 shows the probability that an octahedral site in the fully lithiated composition has a given number of NN V (ranging from 0 to 12). The ground state of $Li_2VO_3$ (blue curve) is a semi-layered structure of Li and $Li/V^{4+}$ layers (56) (shown in Figure S4), in which octahedral Li has 4 and 6 NN V and octahedral V has 3 NN V. For comparison, in a typical layered R-3m $LiMO_2$, such as $LiCoO_2$ or $LiNiO_2$, all the octahedral sites have 6 NN TM. (48) At 1800 K (black), the probabilities of any octahedral site having 3, 4, and 6 NN V decrease, whereas the probabilities of any octahedral site having 1, 2, 5, and 7 NN V sites increase. As the system tends toward the random limit (set to T = $10^5$ K in our simulation), we observe the formation of octahedral sites with 0, 1, and 8 NN V, indicating that disordering increases the Li-rich and metal-rich environments compared with those in the semi-layered $Li_2VO_3$ structure. This result is in line with SRO analysis using Raman spectroscopy (18) that reveal Li-rich and V-rich domains in disordered $Li_2VO_2F$, which is a fluorinated isostructural analogue of $Li_2VO_3$. (26) In the random-limit case (red curve), the probability of the octahedral site environment having a given number of NN V follows an approximately normal distribution with a mean of 4 NN V and a standard deviation of 1.

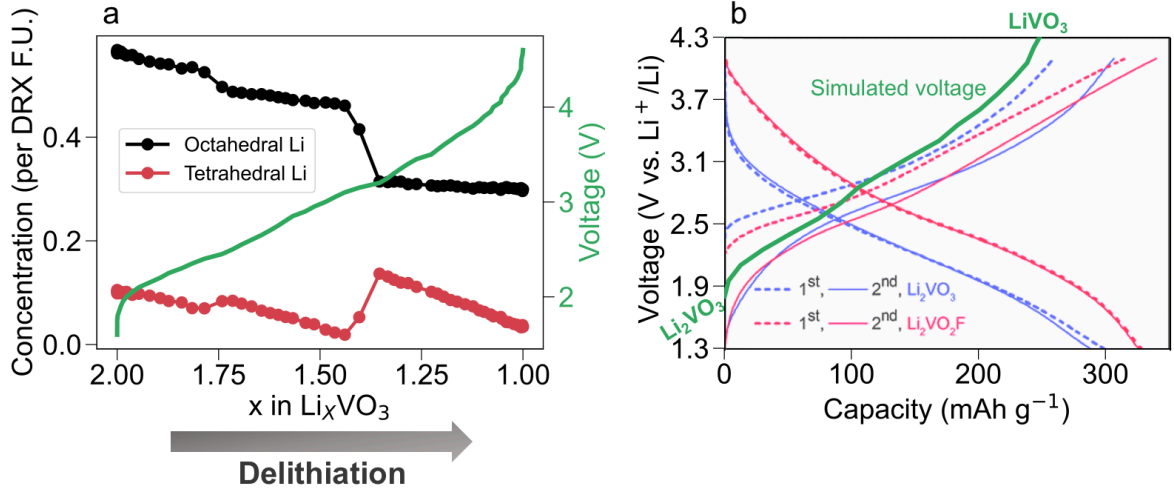

FIG. 3. (a) Concentration of tetrahedral Li (red) and octahedral Li (black) and the voltage profile (green) as a function of Li content in $Li_xVO_3$, 1 < x < 2. The direction of delithiation is indicated by the dark gray arrow under the plot. The Y axis is normalized per DRX F.U. (b) Comparison of experimental voltage profiles (blue curves) obtained by Chen et al. (26) and simulated voltage profile of $Li_2VO_3$ calculated with SGC MC (green curve).

To evaluate the environments from which Li is delithiated, we performed SGC MC simulations and track the concentration of tetrahedral Li and octahedral Li during the delithiation process. Figure 3a presents the voltage profile (green) and the concentrations of tetrahedral Li (red) and octahedral Li (black) for a $Li_xVO_3$ cell with 648 total oxygen atoms for 1 < x < 2. For all compositions, the concentration of octahedral Li is higher than that of tetrahedral Li, consistent with the rocksalt nature of the material. At the fully lithiated composition of $Li_2VO_3$, approximately 10% of all Li are in a tetrahedral site. Although most DRX materials are expected to have only octahedral cations in the fully lithiated state, the presence of tetrahedral Li in $Li_2VO_3$ is not unexpected given that 4% tetrahedral Li has also been observed in the simulated $\sqrt{5} \times \sqrt{5} \times 2$ supercell of $Li_3V_2O_5$ structure. (47) Additionally, the potential for tetrahedral Li formation with varying amount of Li-excess has been demonstrated in the simulation of the fully disordered structure of $Li_{1+y}TM_{1-y}O_2$ (0 < y < 0.3). (48)

Upon delithiation from $X_{Li}$ = 2, there is a gradual decrease in the concentration of both octahedral and tetrahedral Li until approximately $X_{Li}$ = 1.4. At $X_{Li}$ = 1.8, the small increase in the concentration of tetrahedral Li signifies the migration of some octahedral Li into tetrahedral sites. This shift in site occupancies is observed to a much greater extent between $X_{Li}$ = 1.45 and $X_{Li}$ = 1.35 (voltage range of ~3.1–3.2 V). As delithiation proceeds from X = 1.35 to X = 1, the concentration of tetrahedral Li quickly decreases, whereas the concentration of octahedral Li remains almost constant, indicating the preferential extraction of tetrahedral Li at high voltage. At the end of the delithiation ($LiVO_3$), most of the remaining Li is in octahedral sites. Having almost all Li in octahedral sites in $Li_xVO_3$ (x = 1) is consistent with the DFT calculations (Figure S1b), where the configurations with tetrahedral Li are more than 0.1 eV/atom above the hull at $LiVO_3$.

In Figure 3b, the experimental $Li_2VO_3$ voltage profile obtained by Chen et al. (in blue) (26) is compared with our SGC MC calculated profile (in green). The calculated voltage profile has a slightly steeper slope than the experimental profile of the 1st cycle (dotted blue line). (26) Additionally, the average discharge voltage in the experimental profile is reported to be ~ 2.2 V (26), whereas the simulated voltage profile has an average voltage of 2.9 V, calculated using equation 3. This discrepancy between the experimental and theoretical voltage profile is discussed later in this paper.

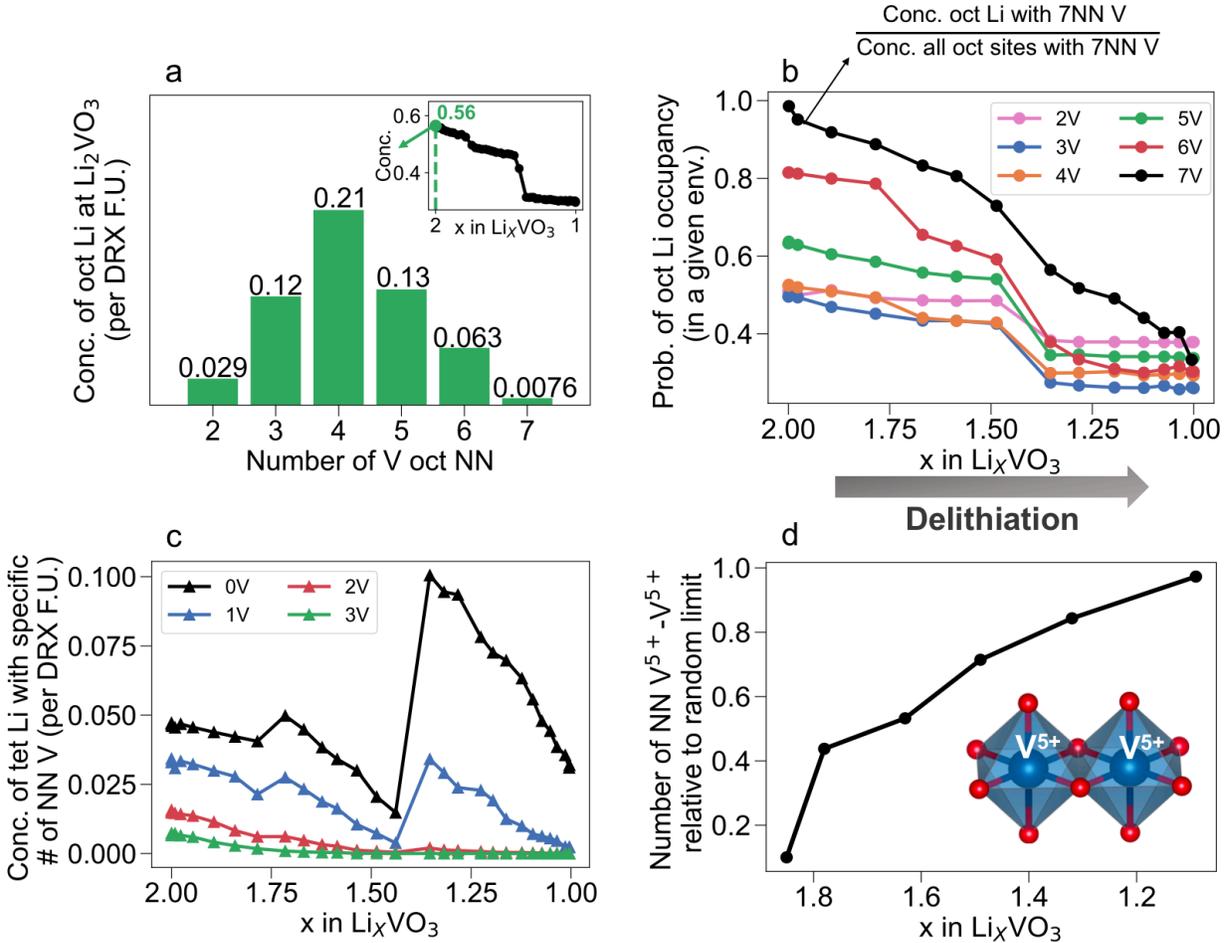

FIG. 4. (a) Concentrations of octahedral Li at $Li_2VO_3$ vs. a certain number of V-occupied NN octahedra. The inset shows the concentration of octahedral Li (black) as a function of Li concentration in $Li_xVO_3$, $1 < x < 2$, and the point $X_{Li} = 2$ is marked with a green filled circle. (b) Probability of octahedral Li occupancy in a given environment as a function of Li content in $Li_xVO_3$, $1 < x < 2$. Octahedral Li sites with 2, 3, 4, 5, 6, and 7 NN V-occupied octahedra are represented by the pink, blue, orange, green, red, and black curves, respectively. The probabilities of octahedral Li sites with 0–1 NN V or > 7 NN V are zero and not shown. The dark gray arrow at the bottom indicates the direction of delithiation. (c) Concentration of tetrahedral Li that face-share with 0–3 V as a function of Li concentration in $Li_xVO_3$, $1 < x < 2$. The black, blue, red, and green curves correspond to 0, 1, 2, and 3 NN V respectively. (d) Number of NN $V^{5+}$–$V^{5+}$ relative to the random limit as a function of Li content.

To better understand how disordered $Li_2VO_3$ is delithiated and how V is oxidized, we analyze the Li and V environments during delithiation in $Li_xVO_3$, as shown in Figure 4. Figure 4a shows the concentrations of octahedral Li before any delithiation at $Li_2VO_3$ as a function of the number of the NN V the Li site has. Altogether, the bars sum to the total concentration of octahedral Li in $Li_2VO_3$, which is 0.56 per F.U. of DRX and is marked in the inset. Most of the Li (1.68 Li in $Li_2VO_3$) prior to delithiation are in octahedral sites, so we focus on octahedral Li in Figure 4a-b. At the fully lithiated composition, octahedral Li with 4 NN V are the most prevalent, whereas those with 7 NN V are the least prevalent.

To understand how Li is removed, we show in Figure 4b the probability (conditional probability) that a site with a certain number of NN V is occupied by Li, as a function of Li content. The total number of octahedral sites with $x$ NN V remains constant as all the V are fixed during topotactic delithiation. The evolution of these curves gives a sense of the environments from which Li is removed at various states of delithiation.

Between $X_{Li} = 2$ and $X_{Li} = 1.45$, the relative contributions to capacity from octahedral sites with 6 and 7 NN V are more significant than those of other environments as their probabilities decrease by more than 0.2, whereas the probabilities of octahedral Li sites with 2–5 NN V decrease by less than 0.1. If there is no preference for how Li is removed (random delithiation), they would be removed at a higher rate from higher frequency environments. Combining Figures 4a and 4b, we see that although the population of octahedral Li with 2 NN V is higher than those with 7 NN V (from Figure 4a), octahedral Li with 2 NN V does not appear to participate in the early stage of delithiation. Thus, while all the octahedral Li environments are extracted to some extent at the beginning of delithiation, there is a clear preference for delithiation of octahedral Li sites with 6 and 7 NN V.

From $X_{Li}$ = 1.45 to $X_{Li}$ = 1.35, the probabilities of all the environments decrease by comparatively similar amounts, indicating that all the environments participate in the delithiation process. This decrease coincides with the rapid growth of tetrahedral Li and removal of octahedral Li, as shown in Figure 3a.

From $X_{Li}$ = 1.35 to $X_{Li}$ = 1, only octahedral Li surrounded by 6 and 7 NN V continue to be removed. This contrasts with the remaining octahedral sites surrounded by 2–5 NN V whose probability of Li occupation barely changes.

From the data in Figure 4b, it is evident that the extraction of octahedral Li is strongly influenced by the number of its NN V. The delithiation preference appears to be in the following order: 7 NN V, 6 NN V, 5 NN V, 4 NNV, 3 NN V, and lastly, 2 NN V environments.

To investigate the environment around tetrahedral Li throughout the delithiation process, in Figure 4c, we track the number of V that face-share with tetrahedral Li in the SGC MC simulations. The quantity plotted is the concentration of tetrahedral Li (per DRX F.U.) with a given number of face-sharing (octahedral) V. The black, blue, red, and green curves correspond to concentrations of tetrahedral Li with 0, 1, 2, and 3 V face-sharing.

Figure 4c shows that for all Li compositions, face-sharing between tetrahedral Li and octahedral V is uncommon. The concentration of tetrahedral Li decreases as the number of face-sharing V increases, consistent with the expected electrostatic penalty. The observation of dilute concentrations of Li-to-TM face-sharing features is not unique to this study and has been observed in the computational studies of overlithiated $Li_{3+x}V_2O_5$ (57) and Mn-based partially-disordered spinel systems. (28,29)

Upon delithiation, from $X_{Li}$ = 2 to $X_{Li}$ = 1.45, the concentration of tetrahedral Li decreases in all environments. Then, at $X_{Li}$ = 1.8, a small increase in environments with 0 and 1 NN V indicates migration of some Li from octahedral sites to tetrahedral sites. This phenomenon is observed to a greater extent near $X_{Li}$ = 1.45, where there is an abrupt increase in the concentration of 0 V and 1 V face-sharing environments. The increase in the tetrahedral Li concentration is consistent with the sudden decrease in the octahedral Li in Figure 3a and will be analyzed further in the discussion section.

To study the spatial arrangements that affect V oxidization, we investigate the total number of NN $V^{5+}$–$V^{5+}$ bonds sampled relative to the random-limit case. To establish these random-limit cases, we sample the $V^{5+}$ ordering during topotactic delithiation occurring at T = $10^5$ K. Our analysis covers Li compositions ranging from $Li_{1.09}VO_3$ to $Li_{1.85}VO_3$, as shown in Figure 4d.

At $X_{Li}$ = 1.85, the number of observed $V^{5+}$–$V^{5+}$ NN pairs is only 20% of what would be present in the random limit. As the concentration of $V^{5+}$ increases, it is no longer possible to avoid the creation of NN $V^{5+}$ environments, and the random limit is reached at $X_{Li}$ = 1 where all V is $V^{5+}$. We reason that the higher electrostatic penalty associated with $V^{5+}$–$V^{5+}$ interactions is responsible for the low occurrence of NN $V^{5+}$-$V^5$ at the early stages of delithiation. This analysis shows that delithiation of Li involves minimizing the number of $V^{5+}$–$V^{5+}$ bonds to minimize the electrostatic interaction between V.

## DISCUSSION

In this work we modelled the Li extraction from a $Li_2VO_3$ DRX compound with cation short-range order. We observed that the local environment of Li generally dictates the order in which it is extracted. Quantification of the Li extraction sequence depends on the frequency with which environments are present (as represented in Figure 4a) and their tendency to hold on to Li (as represented by Figure 4b). The frequency of environments is determined by combinatorial entropy statistics, modified by short-range order, whereas the probably for Li to remain in a specific environment is mostly determined by the interaction of the Li with its environment. Figure 4b reveals that during the delithiation, octahedral Li is preferentially removed from environments with 7 NN V, followed by those with 6 NN V, 5 NN V, 4 NN V, 3 NN V, and ultimately from 2 NN V environments. This delithiation order implies there is a repulsive interaction between Li and V, which is consistent with electrostatics dominating interactions even at this very short-range. Li in 2–5 NN V octahedral environments begin to contribute noticeably to the delithiation process at approximately $X_{Li}$ = 1.45 (around 3.1 V).

Some tetrahedral Li are also extracted from the very beginning of the delithiation process. In particular, the face-sharing Li-V sites appear to deplete faster than the isolated tetrahedral Li.

Critically, we find that the extraction of ~28% of Li ($X_{Li}$ = 1.45) results in a noticeable decrease (increase) in the overall concentration of octahedral (tetrahedral) Li, indicating a migration of Li from octahedral to tetrahedral sites. The same octahedral-to-tetrahedral Li migration has been computationally observed in $Li_{1.44}VO_2F$, a fluorinated isostructural analogue of $Li_2VO_3$. (58) Such collective migration phenomena are known in ordered spinel compounds (28) but they have so far not been well characterized for cation-disordered materials.

As illustrated in Figures 4b and 4c, at the end of delithiation Li is extracted from isolated tetrahedral sites and octahedral sites with 6 and 7 NN V, and the remaining octahedral Li sites with 2–5 NN V appear to be electrochemically inactive.

The order in which specific V ions are oxidized seems to be driven by V–V repulsion. Figure 4d shows that $V^{4+}$/$V^{5+}$ are spatially arranged to minimize $V^{5+}$–$V^{5+}$ bonds until $V^{5+}$–$V^{5+}$ nearest neighbors become unavoidable at the top of charge.

Our study finds a somewhat steeper voltage slope than observed experimentally. Previous modeling studies have demonstrated that the presence of tetrahedral Li can increase the slope of the voltage profile by displacing capacity from low to high voltage. (48) By definition, it has to do so by leaving the average voltage unchanged as this is only determined by the free energies (48) of the $Li_2VO_3$ and $LiVO_3$ endpoints. Our study finds a steeper voltage slope than that observed experimentally, suggesting that the concentration of tetrahedral Li in our model system may be higher than that in the synthesized structure. Our average voltage is also somewhat higher than the measured average voltage for $Li_2VO_3$. (26) Several authors (48,59–61) showed that an increase in disorder can increase the average voltage, suggesting that our theoretical model of the as-synthesized $Li_2VO_3$ endpoint may be slightly more disordered than the experimental reference. (26)

The cation SRO affects the distribution of $Li^+$ migration channels throughout the material and thus the macroscopic diffusivity of Li in $Li_{2-x}VO_3$. (11,26) The experimental Li diffusivity in $Li_2VO_3$ remains mostly constant ($10^{-11}$ $cm^2s^{-1}$) between V = 1.0 V to V = 3.0 V, but rapidly increases to $10^{-10}$ $cm^2s^{-1}$ when charging further to V = ~3.5 V. (26) At higher voltage up to V = ~ 4.2, the diffusivity decreases again to $10^{-12}$ $cm^2s^{-1}$. (26) Figure 3a shows that at $Li_{1.45}VO_3$ (~3.1 V), the concentration of octahedral Li rapidly decreases and is accompanied by an increase in the concentration of tetrahedral Li. This rapid emptying of the octahedral Li into tetrahedral sites could be the origin of the sudden increase in the experimental diffusivity of the $Li_2VO_3$ system from $10^{-11}$ to $10^{-10}$ $cm^2/s$ observed at ~3.5 V. (26) This suggests that the tetrahedral and octahedral occupancy of Li are close in energy at this composition, which would create a dense network of available sites (with similar energy). Although the migration of Li from octahedral to tetrahedral sites can increase the diffusivity at $X_{Li}$ = ~1.45, overstabilization of Li in the tetrahedral sites can also block migration through tetrahedral sites and reduce the diffusivity at the end of the delithiation process. This analysis is in line with the experimental diffusivity profile of this material. (26)

Because the extraction voltage of Li depends strongly on its site energy, understanding which Li is extracted at which voltage provides insight into the local environments to avoid when designing DRX cathode materials. DRX materials generally exhibit voltage profiles that have a large slope due to the various distributions of TMs around the Li sites. (48) This sloping voltage behavior results in reduced capacity within a fixed voltage window, compared to materials with long-range-ordered cations. (48)

It has been theoretically shown that low-voltage TMs, such as V, exhibit steeper voltage slopes when disordered, owing to a weaker screening of Li–TM interactions by oxygen. (48,62) This weaker screening amplifies the effect of cation disorder on the site energy distribution. (48)

For high-voltage TMs (such as Fe, Co, Ni), we expect the number of NN TMs to play a less significant role in the order of delithiation. Their stronger hybridization with oxygen results in more screening which can moderate the effect of different TM environments on the Li site energy. (48) It has indeed been experimentally shown that substituting some of the V with Fe or Ti (63) in $Li_2VO_2F$ can reduce the slope of the voltage, and result in an overall improvement in the electrochemical cyclability.

In addition to the substitution of V with other TMs, optimization of the amount of Li-excess can help to reduce the voltage slope. As discussed earlier, tetrahedral Li is extracted at high voltages of ~4.1 V. If fewer tetrahedral sites are available or if their site energies are raised, the Li in these sites could be extracted at a lower voltage. One potential strategy may be to substitute some of the Li with more TM. Increasing the TM content will decrease the number of 0-TM channels that become high-voltage tetrahedral Li when the surrounding octahedral Li is removed and the channel is occupied by Li. Although decreasing the number of 0-TM in this case can improve the voltage profile, it can be detrimental to the kinetics of Li transport because 0-TM channels can provide low-migration-barrier pathways.

## CONCLUSION

In this work, we investigated the cation short-range order of Li and V in $Li_{2-x}VO_3$. We find that the number of nearest neighbor V in the environment of an octahedral Li plays a crucial role in determining when it is delithiated. We find a strong preference for Li extraction to start from octahedral environments with the most surrounding V. At highly delithiated states, most of the remaining octahedral Li is locked in sites with 2–5 NN V, and Li extraction instead shifts to Li in tetrahedral sites. We find that V oxidizes in a manner that minimizes the number of $V^{5+}$–$V^{5+}$ bonds.

A design implication arising from this study is that future DRX designs should consider the amount of Li-excess required to balance the 0-TM channels needed for percolation with the potential increase in slope resulting from the higher concentration of high-voltage tetrahedral Li. Additionally, over-stabilized tetrahedral Li at the end of the charge can obstruct migration paths and lead to lower $Li^+$ diffusivity and energy density. Surprisingly we also find a partially collective migration of Li from octahedral to tetrahedral sites at partial state of charge, reminiscent of spinels. Although we focused on $Li_2VO_3$ as a model system, our results provide insight for all DRX materials, especially those containing low-voltage TM.

## ACKNOWLEDGMENTS

This work was supported by the Assistant Secretary for Energy Efficiency and Renewable Energy, Vehicle Technologies Office, under the Applied Battery Materials Program, of the U.S. Department of Energy under contract no. DE-AC02-05CH11231. The computational analysis was performed using computational resources sponsored by the Department of Energy's Office of Energy Efficiency and Renewable Energy and located at the National Renewable Energy Laboratory as well as computational resources provided by Extreme Science and Engineering Discovery Environment (XSEDE), supported by the National Science Foundation grant number ACI1053575, and the National Energy Research Scientific Computing Center (NERSC), a DOE Office of Science User Facility supported by the Office of Science and the U.S. Department of Energy under contract no. DE-AC02- 05CH11231. We thank Dr. Daniil A. Kitchaev for the helpful discussion on building cluster expansion of $Li_{2-x}VO_3$. ZJ, LBL and TC acknowledge financial support from the NSF Graduate Research Fellowship Program (GRFP) under contract no. DGE 1752814, DGE 1752814, and DGE 1106400 respectively. Any opinions, findings, conclusions, or recommendations expressed in this material are those of the author(s) and do not necessarily reflect the views of the National Science Foundation.

# Ab-initio study of short-range ordering in vanadium-based disordered rocksalt structures

Zinab Jadidi[1,2], Julia H. Yang[1,2], Tina Chen[1,2], Luis Barroso-Luque[1,2], Gerbrand Ceder[1,2]

[1] Department of Materials Science and Engineering, Berkeley CA 94720
[2] Materials Sciences Division, Lawrence Berkeley National Laboratory, Berkeley CA 94720

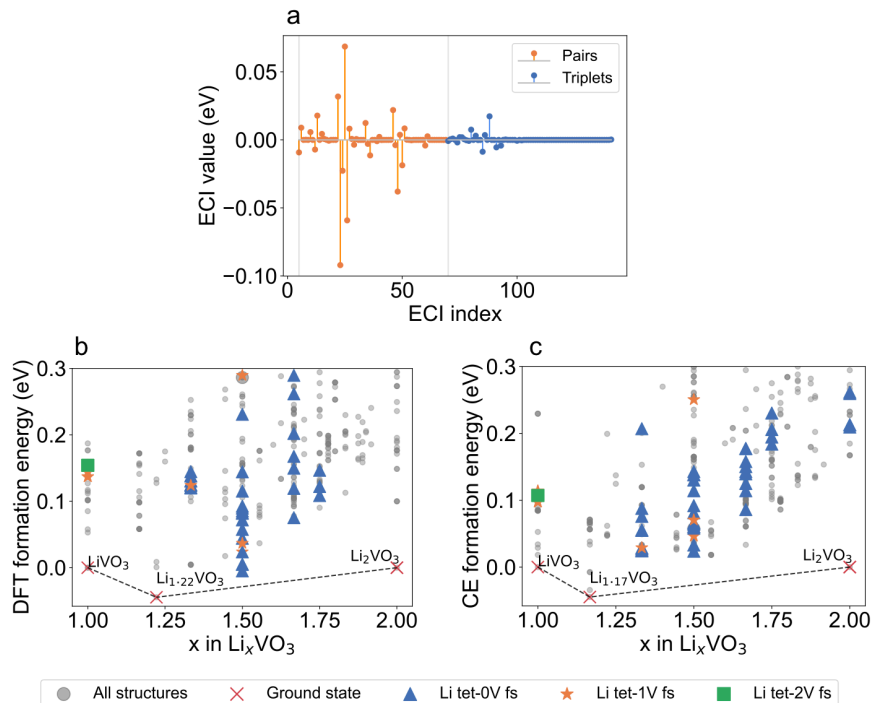

Fig. S1. (a) ECI value vs. ECI index. Each ECI index corresponds to a specific decoration of $Li^+$, $V^{4+}$, and $V^{5+}$ in pair (orange) and triplet (blue) clusters. (b) DFT convex hull showing the energies of structures with an energy above the hull below 0.3 eV. The DFT formation energies of all the structures are marked with gray circles. The ground states ($Li_2VO_3$, $LiVO_3$, and $Li_{1.22}VO_3$) are marked by red crosses. Structures that contain tetrahedral Li with 0, 1, and 2 face-sharing V are marked by blue triangles, orange stars, and green squares, respectively. It is apparent that up to 2 Li tet-V face-sharing is possible based on the DFT data. (c) CE convex hull. The ground states ($Li_2VO_3$, $LiVO_3$, and $Li_{1.17}VO_3$) are marked by red crosses. Although the ground states $LiVO_3$ and $Li_2VO_3$ are the same in both the DFT and CE convex hulls, there is a discrepancy between the DFT and CE convex hull in predicting the ground state for $Li_{1.22}VO_3$. DFT predicts $Li_{1.22}VO_3$ as the ground state, but it is not predicted in the CE convex hull, where $Li_{1.17}VO_3$ is favored instead. Similar discrepancies in preserving the ground states have been observed in cluster expansions of other high-component systems. (1,2)

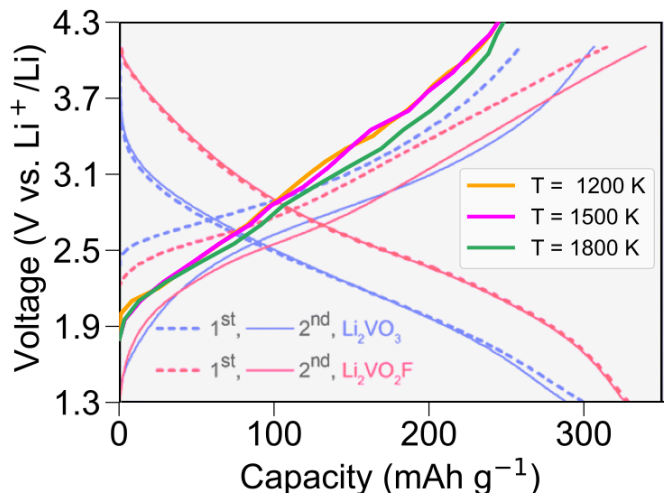

Fig. S2. Comparison between the experimental Li$_2$VO$_3$ voltage profile obtained by Chen et al. in the dotted blue curve (3) with our SGC MC calculated ones in the Li/V framework sampled at T = 1200 K, 1500 K, and 1800 K. The voltage profile obtained at the simulated temperature of 1800 K shows greater similarity to the experimental voltage profile (dotted blue curves) obtained by Chen et al. (3), specifically at higher delithiated states.

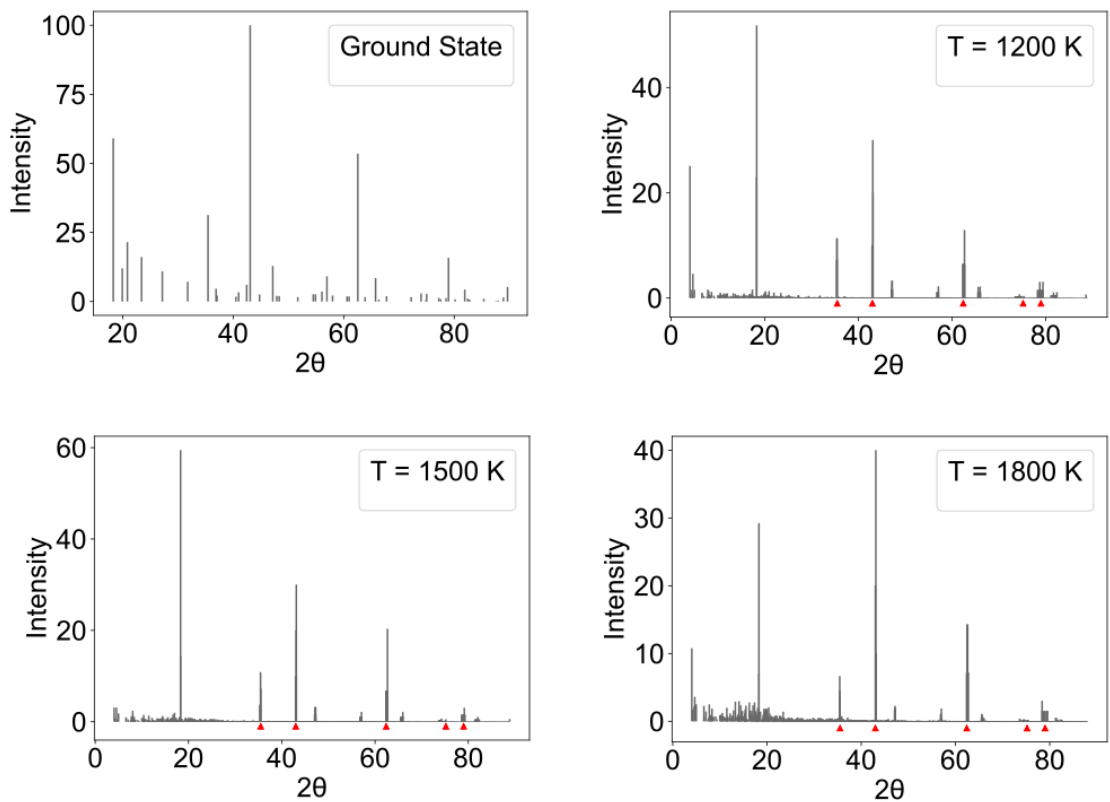

Fig. S3. Simulated XRD pattern of fully lithiated Li$_2$VO$_3$ in 9×8×9 supercell using pymatgen python package. (4) At the ground state, the XRD pattern corresponds to that of the semi-layered ground state of Li$_2$VO$_3$. The simulated XRD patterns for T = 1200, 1500, and 1800 K are averaged over 10 Li/V$^{4+}$ sampled configurations. The main rocksalt peaks (at 38°,43°,63°,76°, and 80°) are marked with red triangles in all cases. While there are peaks at angles below 20 degrees in all three cases, the intensities of the low-angle peaks at 1200 and 1500 K are higher than the main peak at 43°. This, however, does not apply to the Li/V$^{4+}$ configurations sampled at 1800 K.

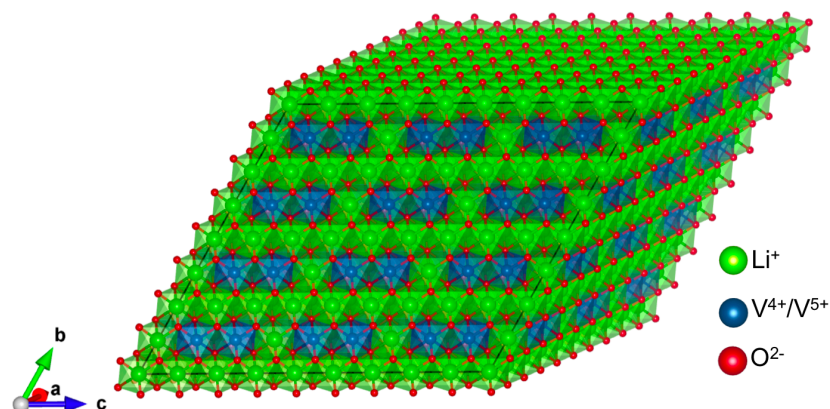

Fig. S4. Schematic of fully lithiated $Li_2VO_3$ at its ground state, which consists of alternating $Li^+$ and $Li^+/V^{4+}$ layers.

### Note 1: Challenges of the CE of the high-component system $Li_{2-x}VO_3$, $0 < x < 1$

This study is based on a lattice cluster expansion of high-component systems with quaternary disorder (considering $Li^+/V^{4+}/V^{5+}$ and vacancies) on octahedral sites and binary disorder ($Li^+$ and vacancies) on tetrahedral sites. The fitting of the ECI for this cluster expansion uses an approach applied in other high-component systems. (1,2,5) In this study, we find there are several challenges associated with structural mapping and sampling when addressing the Li–Vac–V–O system in particular.

First, the $V^{5+}$ octahedron experiences significant distortion in part because it is a $d_0$ element that can strongly distort without an energetic penalty. (6,7) As another example, the presence of vacancies in the system can allow for nearby cations to relax towards the vacancy and reduce electrostatic repulsion from its nearby cations. These distortions make the mapping of the relaxed structure to the lattice-site representation of the cluster expansion extremely challenging as the anion framework becomes highly distorted.

Second, the addition of the interstitial tetrahedral site to the cluster expansion increases the size of the configurational space significantly, and many configurations that could technically exist are practically not achievable. For instance, face-sharing metal-rich clusters have very strong electrostatic repulsion that makes the local configuration unstable. When these configurations are dynamically unstable, they relax away in DFT. As a result, our training data may contain insufficient sampling containing these types of local configurations. We find that this lack of sampling makes the Monte Carlo equilibration more difficult because equilibration may sample structures outside our training set. The dearth of training data with these types of face-sharing configurations may cause our model to represent the energies of structures with face-sharing less accurately and may be a reason for the 2 V– and 3 V–Li tet face-sharing in some of the simulations. Having these face-sharing features as well as the *possibly* larger number of tetrahedral sites compared to the experimental material could be two reasons why the simulated voltage profile is sloppier than the experimental one. Note that metal–Li tet face-sharing features have also been theoretically observed in other studies. (1,5,8)